\documentclass[10pt]{article}

\usepackage{amsmath,epsfig,amssymb,amsfonts,amsthm,epsfig,verbatim,}
\usepackage{color,graphicx,lscape,longtable, mathabx}
\usepackage{natbib}
\usepackage{kotex}
\usepackage{tikz}
\usepackage{multirow}
\usepackage{setspace}
\usepackage{amsthm}
\usepackage{mathtools}
\usepackage[shortlabels]{enumitem}
\usepackage{hyperref}

\newtheorem{Th}{\underline{\bf Theorem}}
\newtheorem{proposition}{Proposition}

\def\bse{\begin{eqnarray*}}
\def\ese{\end{eqnarray*}}
\def\be{\begin{eqnarray}}
\def\ee{\end{eqnarray}}
\def\bsq{\begin{equation*}}
\def\esq{\end{equation*}}
\def\bq{\begin{equation}}
\def\eq{\end{equation}}

\def\Var{\hbox{Var}}

\def\wh{\widehat}

\def\eff{_{\rm eff}}

\def\mR{\mathbb{R}}

\def\sumi{\sum_{i=1}^n}

\def\trans{^{\rm T}}

\def\bmu{\boldsymbol\mu}

\def  \bxi{\boldsymbol\xi}

\def\bb{{\boldsymbol{\beta}}}

\def\0{{\bf 0}}
\def\A{{\bf A}}

\def\B{{\bf B}}
\def\c{{\bf c}}

\def\g{{\bf g}}

\def\h{{\bf h}}

\def\M{{\bf M}}
\def\m{{\bf m}}

\def\S{{\bf S}}

\def\u{{\bf u}}
\def\m{{\bf m}}

\def\u{{\bf u}}

\def\X{{\bf X}}
\def\S{{\bf S}}
\def\x{{\bf x}}

\def\bq{\begin{equation}}
\def\eq{\end{equation}}

\def\wh{\widehat}

\def\trans{^{\rm T}}

\def\squarebox#1{\hbox to #1{\hfill\vbox to #1{\vfill}}}
\def\btheta{{\boldsymbol \theta}}

\def\mH{\mathcal{H}}

\def\bse{\begin{eqnarray*}}
\def\ese{\end{eqnarray*}}
\def\be{\begin{eqnarray}}
\def\ee{\end{eqnarray}}
\def\bsq{\begin{equation*}}
\def\esq{\end{equation*}}
\def\bq{\begin{equation}}
\def\eq{\end{equation}}
\def\wh{\widehat}

\def\trans{^{\rm T}}

\def\boxit#1{\vbox{\hrule\hbox{\vrule\kern6pt\vbox{\kern6pt#1\kern6pt}\kern6pt\vrule}\hrule}}

\newlist{condenum}{enumerate}{1} 
\setlist[condenum]{label=\bfseries Condition \arabic*., 
                   ref=\arabic*, wide}

\def\btheta{{\boldsymbol{\theta}}}

\def\ds{\displaystyle}
\def\S{{\boldsymbol S}}

\begin{document}


\thispagestyle{empty}
\baselineskip 14pt
\renewcommand {\thepage}{}
\include{titre}
\pagenumbering{arabic}
\begin{center}
{\Large\bf Appropriate use of parametric and nonparametric methods in estimating regression models with various shapes of errors}
\end{center}

\begin{center}
Mijeong Kim\\
Department of Statistics, Ewha Womans University, Seoul, 03760, Korea\\
m.kim@ewha.ac.kr\\
\end{center}

\vspace{0.25cm}

\begin{abstract}

In this paper, a practical estimation method for a regression model is proposed using semiparametric efficient score functions applicable to data with various shapes of errors.
 First, I derive semiparametric efficient score vectors for a homoscedastic regression model without any assumptions of errors. 
 Next, the semiparametric eﬃcient score function can be modiﬁed assuming a specific parametric distribution of errors according to the shape of the error distribution or by estimating the error distribution nonparametrically. Nonparametric methods for errors can be used to estimate the parameters of interest or to ﬁnd an appropriate parametric error distribution.
 In this regard, the proposed estimation methods   utilize  both parametric and nonparametric methods for errors appropriately.
Through numerical studies, the performance of the proposed estimation methods is demonstrated.



\end{abstract}

\vspace{0.25cm}

\section{Introduction}

The ordinary least squares (OLS)  is the most commonly used regression estimation under the assumption that errors are not correlated to covariates and are normally distributed with equal variance. However, those assumptions are rather strict; diagnostic methods were made to check the adequacy of the normality assumptions \citep{yazici2007comparison}. By drawing the residual plots, we check whether a specific trend remains that is not interpreted with the assumed model. Q-Q plots can also be used to check whether a normality assumption is valid. When the residuals obtained from the regression analysis are not normally distributed, Box‒Cox transformation is suggested in classical statistics \citep{box1964analysis, spitzer1982primer}. By transforming the dependent variable, the normal error assumption of the regression model becomes valid. However, it has a drawback in that the transformed dependent variable is difficult to interpret. Several researchers have suggested a regression model with errors of a nonnormal distribution. 
\citet{mcdonald1988partially} proposed a partially adaptive estimation for regression models using a generalized $t$ distribution.
\citet{bartolucci2005use} adapt M-estimation for regression models with nonnormal errors using a mixture of normal distributions.
\citet{andersen2008modern} noted that  a robust M-estimator has better efficiency than OLS estimator when errors are not normally distributed.
\citet{cancho2010nonlinear} proposed  a nonlinear regression model with skew-normal errors. 
\citet{usta2011performance} applied  symmetric leptokurtic and skewed leptokurtic distributions for a partially adaptive estimation.
\citet{martin2016semiparametric} proposed a scale mixture model with a nonparametric mixing distribution using a predictive recursion method.
\citet{martinez2017bimodal} postulated a linear regression model with a bimodal distribution on errors. 
\citet{chee2020semiparametric} derived an semiparametric estimation method for the parameters of a linear model with unspeciﬁed symmetric distributed errors. 
 \citet{azzalini2020some} proposed an estimation method for a linear regression with skew-$t$ errors. 
If we want to make a more ﬂexible regression model, we can use a more ﬂexible distribution such as a skewed generalized $t$ distribution  \citep{theodossiou1998financial, davis2015skewed} or
a mixture of two or more distributions for errors.  
However, the more parameters that identify a distribution, the more diﬃcult it is to implement the method numerically. Additionally, even if the computation problem for a more complicated regression model is solved, it is still not possible to represent all error distributions with a ﬁnite number of parameters, so in that case, it would be better to consider a nonparametric method for error distribution.

To coin a ﬂexible regression estimation method for data with various shapes of errors, we set a semiparametric regression model without assuming a speciﬁc distribution for the errors. 
The variance of the semiparametric estimator of \citet{tsiatis2006semiparametric}  asymptotically equals the eﬃciency bound, also known as the Cram\'er-Rao lower bound.
After obtaining semiparametric efficient score functions following \citet{tsiatis2006semiparametric}, we can modify the error-related terms with a speciﬁc distribution or apply nonparametric methods such as a kernel density estimation for errors. 
In this paper, the goal is to suggest a practical method to analyze regression models with unknown error distributions in both parametric and nonparametric ways. 
In Section 2,  I derive semiparametric efficient score vectors for a homoscedastic regression model according to \citet{tsiatis2006semiparametric}.
In Section 3, I propose how to modify the semiparametric eﬃcient score vectors ﬁtted for various shapes of error distributions using both parametric and nonparametric methods. 
In Section 4, simulations and a real data example are provided to show the performance of the proposed method. 
In Section 5, I summarize the study.

\section{Semiparametric regression models} \label{sec:models}

I consider the following  regression model with an error of mean zero and equal variance $ v$.
\be \label{eq:model}
Y=m(\X;\bb)+\epsilon,~~ E(\epsilon)=0, ~~ \Var(\epsilon)= v,
\ee
where $Y \in \mR$ is the response variable, and
$\X \in \mR^l$ is a covariate vector.
The mean function $m$ is a known linear or nonlinear function with the unknown parameter vector $\bb \in \mR^k$.
The goal is to estimate $\btheta=(\bb\trans, v)\trans$. The vector $\btheta$ is $q=k+1$ dimensional.
I assume that covariates and errors are independent and do not impose any distribution assumption on $\epsilon$. 
In this respect, (\ref{eq:model}) is a semiparametric model with parameter of interest $\btheta$ and a nonparametric part associated with $\epsilon$.
Since the distribution of $\epsilon$ is unspecified, it can be said that the distribution of $\epsilon$ is infinite-dimensional nuisance parameters.
\citet{kim2012efficiency} implemented a semiparametric efficient estimator for the nonlinear regression model when $\epsilon$ and $\X$ are not necessarily independent.
\citet{kim2019semiparametric} showed that the semiparametric efficiency bound is different  under the different assumptions. 
According to \citet{kim2019semiparametric}, a more general estimation method can be used in a special case, but it cannot reach the efficiency bound when the data fit such a special case. 
Thus,  the method of \citet{kim2012efficiency} may not give an efficiency bound for the above homoscedastic error model because the error assumptions of \citet{kim2012efficiency} are different from (\ref{eq:model}).
In addition, the method of  \citet{kim2012efficiency} requires a function that can describe the relationship between  covariates and errors because they are not necessarily independent.
In reality, it is very difficult to assume or estimate  the relationship between them, especially when we have a small number of observations. 
In this paper, I aim to find semiparametric efficient score vectors for (\ref{eq:model}) under the assumption of independence of covariates and errors.

\subsection{Derivation of the semiparametric efficient score function}

The probability density function (pdf) of $(\X,Y)$ is represented as
\bse
f_{\x,y}(\x,y)=f_{\x}(\x)f_\epsilon(y-m(\x;\bb)),
\ese
where $f_{\x}(\x)$ and $f_\epsilon(\epsilon)$ are the pdf of $\X$ and $\epsilon$, respectively.
I consider Hilbert space $\mH$, which includes all mean zero functions with finite variance.
According to \citet{tsiatis2006semiparametric}, the efficient score function is obtained by projecting the score function onto the orthogonal complement space of the nuisance tangent space. 
 In the Hilbert space $\mH$, I first derive the nuisance tangent space $\Lambda$ and its orthogonal complement space $\Lambda^\perp$ to find an efficient score function for 
 $\btheta=(\bb\trans, v)\trans$. 

\begin{proposition} \label{prop:prop1}
The nuisance tangent space $\Lambda$ and its orthogonal complement space $\Lambda^\perp$ are given by
\bse
\Lambda&=&\{\h_1(\x)+h_2(\epsilon): E\{\h_1(\x)\}=\0, E\{ h_2(\epsilon)\}=E\{\epsilon h_2(\epsilon)\}=E\{\epsilon^2 h_2(\epsilon)\}=0\},\\
\Lambda^\perp&=&\{\g(\x,\epsilon):E\{\g(\x,\epsilon)|\X\}=\0, E\{\g(\x,\epsilon)|\epsilon\}=\c_1\epsilon+\c_2t(\epsilon): \c_1,\c_2 \in \mR^{k+1}\},
\ese
where $t(\epsilon)=\epsilon^2- v-E(\epsilon^3)\epsilon/ v$.
\end{proposition}

The proof of Proposition \ref{prop:prop1} is  provided in S2 of Supporting Information.

The projection of any function $\h(\x,\epsilon)\in \mH$  onto $\Lambda^\perp$ is obtained as
\bse
\Pi\{\h(\x,\epsilon)|\Lambda^\perp\}=\h(\X,\epsilon)-E\{\h(\X,\epsilon)|\X\}-E\{\h(\X,\epsilon)|\epsilon\}+\frac{E\{\epsilon\h(\X,\epsilon)\}\epsilon}{ v}
+\frac{E\{t(\epsilon)\h(\X,\epsilon)\}t(\epsilon)}{E[\left\{t(\epsilon)\right\}^2]}.
\ese
Now, we can derive the efficient score vector by projecting score functions of
 $\btheta=(\bb\trans, v)\trans$ on $\Lambda^\perp$.
\begin{Th} \label{th:thm1}
The efficient score vector
$\S\eff(\X,Y;\btheta,f_\epsilon)=\left(\S_{{\rm eff},\bb}\trans (\X,Y;\btheta,f_\epsilon), S_{{\rm eff}, v}(\X,Y;\btheta,f_\epsilon)\right)\trans $ is given by
\be
\S_{\rm{eff},\bb} (\X,Y;\btheta,f_\epsilon)
&=&-\frac{f'_\epsilon(\epsilon)}{f_\epsilon(\epsilon)}
\left[ \m'_{\bb}(\X,\bb)-E\left\{\m'_{\bb}(\X,\bb)\right\}\right]
+\left( \frac{\epsilon}{ v}-\frac{E(\epsilon^3)t(\epsilon)}
{ v E[\left\{t(\epsilon)\right\}^2] }\right)
E\left\{\m'_{\bb}(\X,\bb)\right\}      \nonumber \\
S_{\rm{eff}, v} (\X,Y;\btheta,f_\epsilon)
&=&\frac{t(\epsilon)}{E[\left\{t(\epsilon)\right\}^2] }.  \label{eq:seff}
\ee

\end{Th}

The proof of Theorem \ref{th:thm1} is  provided in S3 of Supporting Information.

The estimates of $\btheta=(\bb\trans, v)\trans$ can be obtained by solving $\sumi \S\eff(\X_i,Y_i;\btheta,f_\epsilon)=\0$.
Details of the asymptotic property of  the proposed semiparametric estimator following \citet{tsiatis2006semiparametric}  are described in S1 of Supporting Information.
In particular, if $\epsilon$ is distributed as $N(0, v)$, then  we can plug
$\ds{\frac{f'_\epsilon(\epsilon)}{f_\epsilon(\epsilon)}}=-\frac{\epsilon}{ v}$,
and $E[\left\{t(\epsilon)\right\}^2]=2 v^2$ into the above equations.
Then, it follows that
\bse
\S_{{\rm eff},\bb} (\X,Y;\btheta,f_\epsilon) 
=\frac{\epsilon}{ v} \m'_{\bb}(\X,\bb),~~
S_{{\rm eff}, v} (\X,Y;\btheta,f_\epsilon) 
=\frac{\epsilon^2- v}{2 v^2}.
\ese

In the case of linear regression with $m(\X,\bb)=\X\trans\bb$, we obtain the  estimator for $\bb$ by solving
\bse
\sumi \S_{{\rm eff},\bb} (\X,Y;\btheta,f_\epsilon) =\frac{1}{ v}\sumi\epsilon_i\X_i=\frac{1}{ v}\sumi(Y_i-\X_i\trans\bb)\X_i=\0.
\ese
The estimator obtained by solving the above equation is equal to the OLS estimator.

\subsection{Efficiency bound}\label{sec:efficiency}

According to Theorem 4.1 in  \citet{tsiatis2006semiparametric}, 
 the asymptotic variance of the estimator $\wh\btheta$ is given by 
\be \label{eq:estimated_cov}
\mbox{cov}(\wh\btheta)=n^{-1}\left[E\left\{\S\eff(\X,Y;\btheta_0,f_\epsilon)\S\eff(\X,Y;\btheta_0,f_\epsilon)\trans \right\}\right]^{-1}.
\ee
Let $\M_1$ be the inverse matrix of the semiparametric efficiency bound, that is,
$\M_1=E\left\{\S\eff(\X,Y;\btheta_0,f_\epsilon)\S\eff(\X,Y;\btheta_0,f_\epsilon)\trans \right\}.$
The proposed efficient score function vector (\ref{eq:seff}) has a different form from Theorem 1 of  \citet{kim2012efficiency}. Note that \citet{kim2012efficiency} do not assume that $\epsilon$ and $\X$ are necessarily independent.
 Under the assumption that covariates and errors are independent, the semiparametric efficient vector of \citet{kim2012efficiency} takes  the following form.
\be
\S_{\rm{eff,   2}}(\X,Y;\btheta,f_\epsilon)=
\begin{pmatrix}
\ds{\m'_{\bb}(\X,\bb)\left\{ \frac{\epsilon}{ v}-\frac{E(\epsilon^3) t(\epsilon)}{ v E[\left\{t(\epsilon)\right\}^2]} \right\}} \\
\ds{\frac{t(\epsilon)}{E[\left\{t(\epsilon)\right\}^2] }} \label{eq:type2_Seff}
\end{pmatrix}.
\ee
In this case, the semiparametric efficiency bound is equal to $\M_2^{-1}=\left[E\left\{\S_{\rm{eff,   2}}(\X,Y;\btheta_0,f_\epsilon)\S_{\rm{eff,   2}}f(\X,Y;\btheta_0,f_\epsilon)\trans \right\}\right]^{-1}$.

Here, we can show that the semiparametric efficiency bounds are different under the different assumptions, as \citet{kim2019semiparametric} verified.
After some calculations, we  have $\M_1-\M_2=E(\u\u\trans)$, a nonnegative definite under the assumption that $\epsilon$ and covariates are independent, where
\bse
\u=\begin{pmatrix}
\left(\frac{f'_\epsilon(\epsilon)}{f_\epsilon(\epsilon)}+ \frac{\epsilon}{ v}-\frac{E(\epsilon^3)t(\epsilon)}
{ v E[\left\{t(\epsilon)\right\}^2] }\right)
\left[ \m'_{\bb}(\X,\bb)-E\left\{\m'_{\bb}(\X,\bb)\right\}\right] & \0 \\
\0 & \0
\end{pmatrix}.
\ese
This implies that the proposed estimator has smaller variance for $\bb$ than using  the above semiparametric efficient scores (\ref{eq:type2_Seff}),
and both efficient score functions for $ v$ are equal.

\section{Estimation} \label{sec:estimation}


We can obtain a semiparametric efficient estimator by solving (\ref{eq:seff}).
However, there are unknown terms such as $f_\epsilon(\epsilon)$ and $E(\epsilon^3)$ in  (\ref{eq:seff}).
When those terms are estimated properly, the estimator that is very close to the true parameter will be obtained.
To substitute an error distribution $f_\epsilon(\epsilon)$, we can approach it in two ways, a parametric and a nonparametric method.

\subsection{Nonparametric methods for errors} \label{sec:nonpara}
We can consider a kernel density estimation, logspline density estimation \citep{kooperberg1991study} and log-concave density estimation \citep{rufibach2007computing} as a nonparametric approach.
Because we need the first derivative function for the pdf of $\epsilon$ in (\ref{eq:seff}), the following kernel density estimation would be more appropriate.
\be
\wh f_\epsilon(\epsilon;h)=\frac{1}{nh}\sumi K\left(\frac{\epsilon-\wh\epsilon_i}{h}\right), \label{eq:kernel} ~~~~
\wh f_\epsilon'(\epsilon;h)=\frac{1}{nh^2}\sumi K'\left(\frac{\epsilon-\wh\epsilon_i}{h}\right), \nonumber
\ee
where $K$ is a kernel function and $h$ is the bandwidth. 
Following Chapter 2.5 of \citet{wand1994kernel}, we assume the same  regularity conditions, which are described in S4 of Supporting Information.
Through the study, I use the Gaussian kernel and the optimal bandwidth $h=\{{4\wh\sigma^5}/{(3n)}\}^{1/5}\approx 1.06\wh\sigma n^{-1/5}$, where $\wh\sigma$ is the sample standard deviation and $n$ is the sample size.  
It is known that the optimal  bandwidth minimizes the mean integrated squared error \citep{silverman1986density}.
We can esimate the third and fourth moments as
\bse
\wh E(\epsilon^3)=\frac{1}{n}\sumi\wh\epsilon_i^3,~~ \wh E(\epsilon^4)=\frac{1}{n}\sumi\wh\epsilon_i^4.
\ese

\begin{Th} \label{th:thm3}
Assume $E\left\{ \S\eff(\X,Y;\btheta, f_0)\right\}=\0$ has a unique root and $\wh\btheta$ satisfies that
\bse
\sumi \S\eff(\X_i,Y_i;\btheta, \wh f_\epsilon(\epsilon;h))=\0,
\ese
Then under the regularity condition, $\wh\btheta$ satisfies 
\be \label{eq:asymptotic_nonpara}
n^{1/2} (\wh\btheta-\btheta_0)\to N\left( \0, \left[E\left\{ \S\eff(\X,Y;\btheta, f_0)^{\otimes 2}\right\}\right]^{-1}\right).
\ee
in distribution as $n\to\infty$.
\end{Th}

The proof of Theorem \ref{th:thm3} is  provided in S4 of Supporting Information.

If we have a sufficient number of observations, we can approach to estimate errors in a nonparametric way. Otherwise, it would be better to find an appropriate parametric distribution for error by repeating trial and error. Even if we do not have enough samples, checking residuals obtained from a nonparametric method will be helpful to find appropriate parametric distributions for errors.
Since the regression model includes parameters of interests and errors are estimated by a nonparametric method,  this estimation method  will be hereinafter referred to the semiparametric estimation method.

\subsection{Parametric methods for errors}
Once we try the OLS method, we test the validity of the normal assumption for errors with various methods, such as the Shapiro‒Wilk test and checking the pattern of residuals. If residuals have normal patterns, we can stop there and report the OLS estimator. Otherwise, we need to try to assume another distribution. In this case, we can try semiparametric efficient estimation with kernel density estimation for errors. When the residual pattern is unimodal,  
the Cullen and Frey graph \citep{cullen1999probabilistic} helps to choose a feasible distribution.
Cullen and Frey graphs show the sample skewness and sample kurtosis
In \textsf{R}, the package \texttt{fitdistrplus}  function \texttt{desc} was implemented \citep{delignette2015fitdistrplus}.
If the residual is symmetric and unimodal, various symmetric distributions, such as logistic and $t$ distributions, can be good candidates. 
When $\epsilon$ is skewed and unimodal, we can use skewed distributions for $f_\epsilon(\epsilon)$, such as $\chi^2$ distributions and Gumbel distributions.
If $\epsilon$ has a bimodal pattern, a mixture of two normal distributions can be applicable  for $f_\epsilon(\epsilon)$.
According to the used distribution of $\epsilon$, we can modify the  semiparametric efficient score function (\ref{eq:seff}).
Note that
  we need to check the pattern of residuals such as the Q-Q plot and histogram after the estimation procedure. We can find an appropriate distribution by repeated trials. The following are some examples of parametric methods for errors:

{\bf Example 1}  \,\,\, Minimum extreme value distribution  (Gumbel  distribution)
\be
f_\epsilon(\epsilon;\lambda)=\frac{1}{\lambda}\exp\left\{ \frac{\epsilon-\lambda\gamma}{\lambda}-\exp\left(\frac{\epsilon-\lambda\gamma}{\lambda}\right)\right\}, \label{eq:gumbel}
\ee
where $\lambda>0$ and $\gamma$ is the Euler–Mascheroni constant, close to 0.5772.
The minimum extreme value distribution is left skewed.
Do not confuse with the maximum extreme value distribution, which is also called the Gumbel distribution but right-skewed.
We have 
\bse
\frac{f'_\epsilon(\epsilon;\lambda)}{f_\epsilon(\epsilon;\lambda)}=\frac{1}{\lambda}-\frac{1}{\lambda}\exp\left(\frac{\epsilon-\lambda\gamma}{\lambda}\right),
~~E(\epsilon)=0,~~E(\epsilon^3)=-2\lambda^3\zeta(3), ~~ E(\epsilon^4)=\frac{\lambda^4\pi^4}{15},
\ese
where $\zeta(\cdot)$ is the Riemann zeta function.
We plug the above terms into  (\ref{eq:seff}), and then we obtain an efficient score function corresponding to the error of the Gumbel distribution.

{\bf Example 2}  \,\,\,  Mixture of two normal distributions
\bse
f_e(e;p_1,p_2,m_1,m_2,\sigma_1,\sigma_2)= \sum_{i=1}^2p_if_{e,i}(e;m_i,\sigma_i),  
\ese
where $\ds{f_{e,i}(e;m_i,\sigma_i)=\frac{1}{\sigma_i}\phi\left(\frac{e-m_i}{\sigma_i} \right)}$ for $i=1,2$, $p_1=1-p_2$ and $\phi$ is a pdf of a standard normal density.
Then, we have  $\epsilon=e-m_0$, which satisfies  that $E(\epsilon)=0$, where $m_0=\sum_{i=1}^2 p_i m_i$. 
Let  $\bmu\trans=(p_1,p_2,m_0,m_1,m_2,\sigma_1,\sigma_2)\trans$ and $   \bmu_i\trans=(m_0,m_i,\sigma_i)\trans$ for $i=1,2$. 
We denote  $\ds{f_{\epsilon,i}(\epsilon;   \bmu_i)=\frac{1}{\sigma_i} \phi\left(\frac{\epsilon-m_i+m_0}{\sigma_i} \right)}$.
Then, it follows that
\be
f_\epsilon(\epsilon;   \bmu)&=&\sum_{i=1}^2p_if_{\epsilon,i}(\epsilon;   \bmu_i),~~
\frac{f'_\epsilon(\epsilon;   \bmu)}{f_\epsilon(\epsilon;   \bmu)}=\frac{ \sum_{i=1}^2p_if'_{\epsilon,i}(\epsilon;   \bmu_i)}{ \sum_{i=1}^2p_if_{\epsilon,i}(\epsilon;   \bmu_i)},  \label{eq:gmixture} \\
E(\epsilon^3)&=&\sum_{i=1}^2p_i\{(m_i-m_0)^3+3(m_i-m_0)\sigma_i^2\}, \nonumber \\
E(\epsilon^4)&=&\sum_{i=1}^2p_i\left\{(m_i-m_0)^4+6(m_i-m_0)^2\sigma_i^2+3\sigma_i^4\right\}, \nonumber
\ee
where $f_i'(\epsilon;   \bmu)=-\left(\frac{\epsilon-m_i+m_0}{\sigma_i^2}\right) f_i(\epsilon;   \bmu)$.
Plugging the above terms into (\ref{eq:seff}), we obtain the efficient score function corresponding to the regression model with errors of mixed two normal distributions.

As we have seen in Example 1 and Example 2, we need to estimate additional parameters such as $\lambda$ and $\bmu$ to obtain the parameters of interest $\wh\btheta$.
Maximum likelihood estimators  can be obtained for additional parameters by solving corresponding score functions.
In Section \ref{sec:efficiency}, I derived the efficiency bound $\M_1^{-1}$ of the semiparametric efficient estimator  for (\ref{eq:model}).
Even if we select a density included in the same distribution family as the true density of $\epsilon$, we need to estimate additional parameters such as a scale parameter that determines  the specific form of the density.
Let the additional parameter vector be $\bxi$. By estimating $\btheta$ using the estimate of additional parameter $\bxi$, we may have a different covariance matrix from $n^{-1}\M_1^{-1}$ in  (\ref{eq:estimated_cov}).
The estimated covariance matrix can be derived when using an additional parameter $\bxi$ in the following way.

\begin{Th} \label{th:thm2}
Assume the true density of $\epsilon$ is $f_0(\epsilon)=f_\epsilon(\epsilon;   \bxi_0)$ and
\bse
E\left\{ \S\eff(\X,Y;\btheta, f_\epsilon(\epsilon, \bxi)\right\}=\0,~~ E\{\S_{   \bxi}(\X,Y;\btheta, f_\epsilon(\epsilon,  \bxi))\}=\0
\ese
has a unique root. The parameter estimate of $   \bxi$ is obtained as $\wh   \bxi$.
Let
\bse
\A=E\left\{ \frac{\partial \S\eff(\X,Y;\btheta, f_\epsilon(\epsilon, \wh   \bxi))}{\partial \btheta\trans}\right\}~~\mbox{and~~}
\B=E\left\{ \S\eff(\X,Y;\btheta, f_\epsilon(\epsilon, \wh   \bxi))^{\otimes 2}\right\}
\ese
be bounded and nonsingular matrices.
Then the estimator $\wh\btheta$ that is obtained by solving
\bse
\sumi  \S\eff(\X_i,Y_i;\btheta, f_\epsilon(\epsilon,    \bxi))=\0,~~ \sumi \S_{\bxi}(\X_i,Y_i;\btheta, f_\epsilon(\epsilon,    \bxi))=\0
\ese
satisfies
\be \label{eq:asymptotic_addpara}
n^{1/2} (\wh\btheta-\btheta_0)\to N\left\{ \0, \A^{-1}\B(\A^{-1})\trans\right\}.
\ee
in distribution as $n\to\infty$.
\end{Th}

Theorem \ref{th:thm2} can be easily proven by Taylor expansion and I omit it.

Although a semiparametric model was used to derive the efficient score function, a parametric distribution was used in the error estimation procedure.
Thus, this method will be hereinafter referred to as a parametric estimation to prevent confusion.

\section{Numerical studies} \label{sec:simu}

\subsection{Simulations}

\subsubsection{Nonlinear regression} \label{sec:simu_nonlinear_regression}
In this subsection, simulations were conducted for a nonlinear regression model with various types of errors to  show the finite sample performance of the proposed method.
For a nonlinear mean function $m$ in (\ref{eq:model}), an exponential model was used as follows.
\bse
m(X;\bb)=\beta_1\exp(\beta_2X),
\ese
where $\bb=(\beta_1,\beta_2)\trans=(12,-0.5)\trans$.   Covariate $X$  is drawn from Gamma$(2.5,1.5)$, and the model error $\epsilon$ was generated in two following different settings: 
skewed unimodal errors and  bimodal errors. 
\begin{enumerate}[(a)] 
\item Gumbel distribution with $\lambda=1.5$ in  (\ref{eq:gumbel}).
\item  Gaussian mixture with  $\bmu\trans=(p_1,p_2,m_0,m_1,m_2,\sigma_1,\sigma_2)\trans=(0.6,0.4,0,-2,3,0.6,0.7)\trans$  in (\ref{eq:gmixture}).
\end{enumerate}

Then, the variance $ v$ of Simulations (a) and (b) are obtained as 3.7011 and 6.4120, respectively.
One thousand simulations were conducted for sample sizes of $n=200, 300, 500$ and $1000$.
To estimate the parameter $\bb$, we need to find proper efficient score functions, as explained in Section \ref{sec:estimation}.
We can use a parametric method by plugging components corresponding to an assumed distribution into (\ref{eq:seff}).
In a parametric approach, an assumed distribution plays an important role in estimating the parameters of interest. 
For the parametric approach, the true distribution of  $\epsilon$ and normal distribution were used to find the  efficient score functions.
To evaluate the performance of the proposed method, I also report a semiparametric estimation method  that is explained in subsection \ref{sec:nonpara}.

The results  for Simulations (a) and (b) are presented in  \ref{simu2} and \ref{simu3}, respectively.
To verify asymptotic properties  (\ref{eq:asymptotic_nonpara}) and (\ref{eq:asymptotic_addpara}) of the proposed methods, we need to compare the standard error obtained from estimators with the estimated standard errors to see if they are close.
I reported median of estimates of parameters among 1000 simulations, the standard error of 1000 estimates and median of estimated standard error among 1000 simulations,
which are represented as Estimate, SE1 and SE2, respectively, in Tables  \ref{simu2} and \ref{simu3}.
The value corresponding to the smallest SE1 among those obtained from the three methods is indicated in bold. 
In each table,  95\%cvg represents the coverage that the 95\% confidence interval includes the true parameter value. 
Assuming normal errors, we can use the Shapiro‒Wilk test to check whether the residuals violate the normal assumption. 
The Shapiro‒Wilk test results are given in Table \ref{table:shapiro}, in which the number represents the percentage of cases where the residuals violate the normality assumption.

Table \ref{simu2} presents the results of Simulation (a). When the parametric method with true error density is used,  $\wh\bb$ has the smallest standard error and 95\%cvg also shows a better result than using other methods.
In the case of $n=200$ of the parametric method with true error density, it is difficult to say that SE1 and SE2 of $\wh\beta_1$ are close, but as $n$ increases, the difference narrows.
In terms of 95\%cvg,  it shows somewhat unstable results when $n=200$.
The Shapiro‒Wilk test results in Table  \ref{table:shapiro}  conclude that the normal density assumption for errors is not appropriate for the Gumbel errors. 
The semiparametric estimation method overall performs better than the parametric method with normal pdf.
Table \ref{simu3} shows the results of Simulation (b). 
In terms of estimator variability, the parametric method with the true pdf gives the best results.
In all cases, when using the parametric method with normal pdf, SE1 is much larger than when using the parametric method with the true pdf.
The Shapiro‒Wilk test in Table  \ref{table:shapiro}  also clearly shows that the normal error assumption is not appropriate. 
On the other hand, when the semiparametric estimation method is used, SE1 is not significantly different than that of the parametric method with true pdf.

\subsubsection{Linear regression with skew-$t$ errors}
In this subsection, we conducted simulations  for  a linear regression with errors of  a more flexible distribution.
A skew-$t$ distribution is a well-known flexible distribution that is identified by four parameters: location $\xi$, scale $\omega (>0)$, shape (or slant) $\alpha$, and tail-weight $\nu (>0)$ parameters \citep{azzalini2020some}.
Its standard (location $\xi=0$ and scale $\omega=1$) univariate pdf is given by
\be 
t(z;\alpha,\nu)=2t_1(z;\nu)T_1\left( \alpha z \sqrt{\frac{\nu+1}{\nu+z^2}} ;\nu+1    \right), ~~z\in\mR, \label{eq:skewt}
\ee
where
$t_1(z;\nu)$  is a classical Student's $t$ with $\nu$ degrees of freedom and $T_1(\cdot;\nu+1)$ is the cumulative distribution function (cdf) of $t_1(\cdot;\nu+1)$.
A transformation of (\ref{eq:skewt})  is given by $Y=\xi+\omega Z$, then, its pdf becomes
\bse
t_Y(y;\xi,\omega,\alpha,\nu)=\omega^{-1} t(z;\alpha,\nu), ~~ z=\omega^{-1}(y-\xi).
\ese
It is written that $Y\sim ST(\xi,\omega^2,\alpha,\nu)$.
Although a skew-$t$ distribution with four parameters ensures model flexibility, \citet{azzalini2020some} note that computational difficulties arise when finding the roots of the distribution parameters as the number of parameters increases.
Similarly, computational issues are also prone to occur when using the proposed parametric method with an error density that includes many parameters.
Thus, in this case, the proposed semiparametric estimation method is useful to avoid estimating many parameters of a skewed error distribution.
I compare the proposed semiparametric estimation method and the parametric method suggested in \citet{azzalini2020some}.

The simulations were implemented in the following ways.
A linear mean function $m$ in (\ref{eq:model}) is given by
\bse
m(X;\bb)=\beta_0+\beta_1X_1+\beta_2X_2,
\ese
where $\bb=(\beta_0,\beta_1,\beta_2)\trans=(5,1,1.8)\trans$.  
\begin{enumerate}[(a)] 
\item Generate $n=300$ samples of $\epsilon$ from $ST(-2.46,3^2,2.5,10)$.
\item Among $n=300$ samples of $\epsilon$, generate 70\%  from $ST(-2.46,3^2,2.5,10)$
and 30\%  from the distribution obtained by location transforming the  Gamma$(2.5,3)$ by 7.5 to the left.
\end{enumerate}

For both simulations, 1000 iterations were performed.
The \textsf{R} package \texttt{sn} \citep{azzalini2022package} provides a function \texttt{rst} that generates random numbers of skew-$t$ distribution
and a function \texttt{selm} that implements maximum likelihood estimation (MLE) for a linear regression with skew-$t$ errors following \citet{azzalini2020some}.
In Simulation (a), the data have errors that are generated from the skew-$t$ distribution. 
I used  \texttt{selm} for the parametric method and the proposed semiparametric method incorporating kernel density estimation to compare the variability of the estimated parameters of a linear model.
In Simulation (b), errors were generated from a mixture of skew-$t$ and gamma distributions. 
For simplicity, I denote this distribution as a perturbed skew-$t$ distribution. 
It could be one case we encounter in real life. 
We do not know exactly what the error distribution of the data is, but we often obtain such data with skewed errors. 
We can consider a skewed generalized $t$ distribution (SGT), which is a highly flexible distribution identified by five parameters \citep{theodossiou1998financial, davis2015skewed}.
However,  the algorithm for the estimation method for the regression model with SGT errors has not been implemented to date.
For this reason, it would be reasonable to use \texttt{selm} when performing a regression analysis on data with skewed errors.
Even from the histogram of the data, we can find a similar pattern as SGT, so it seems natural to use \texttt{selm}.
In addition, the proposed semiparametric estimation method can be applied without assuming any specific form of the error distribution.

\begin{figure}
\includegraphics[scale=0.75]{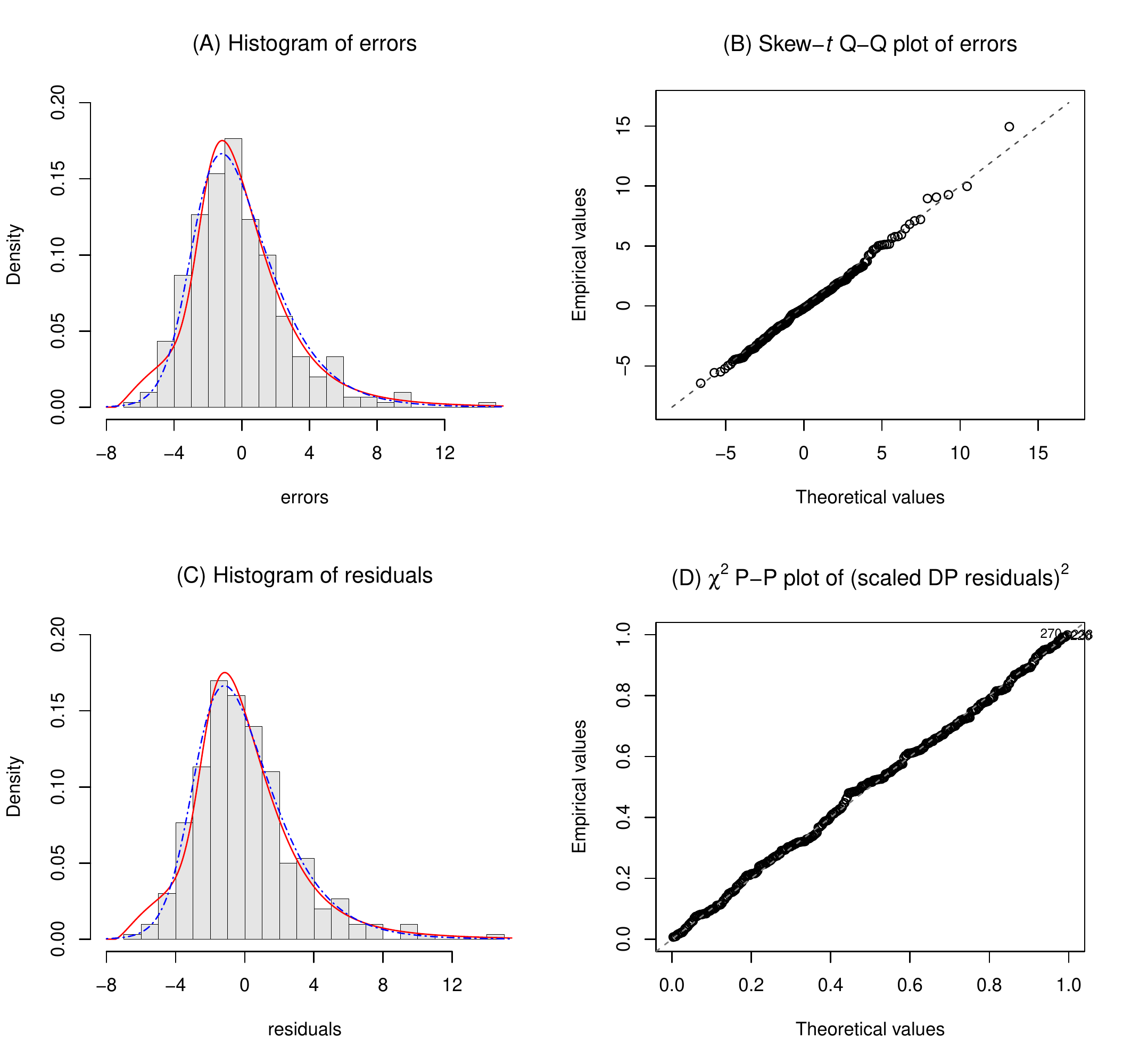}
\caption{Diagnostics of  the parametric estimation using  \texttt{selm}  for one case of Simulation (b). 
(A) Histograms of errors generated from a perturbed  skew-$t$ distribution. 
The red curve and  the blue dashed curve represent the true distribution of errors and the fitted skew-$t$ distribution using \texttt{selm}, respectively. 
(B) Skew-$t$ Q-Q plot of errors generated from a perturbed  skew-$t$ distribution. 
(C) Histograms of residuals obtained from the result using  \texttt{selm}. 
The red curve and  the blue dashed curve represent the true distribution of errors and the fitted skew-$t$ distribution using \texttt{selm}, respectively. 
(D) $\chi^2$ P-P plot of squares of scaled DP residuals. \label{figure1} }
\end{figure}

In Table \ref{table:SELM},   the results of  \texttt{selm} and the proposed method are represented.
In the proposed method,  the OLS estimator was used for the initial values.
Estimate, SE1, SE2 and 95\%cvg of Table \ref{table:SELM} indicate the median of estimates of parameters, the empirical standard error, median of estimated standard error and the coverage of the 95\% confidence interval, respectively.
The smaller value of SE1 is shown in bold between the empirical standard errors of the two methods. 
In Simulation (a), both methods provide consistent estimators.
In terms of the estimator variability, the parametric method using \texttt{selm}  performs better than the semiparametric estimation method.
When the error distribution exactly follows skew-$t$, the method of \citet{azzalini2020some} results in more efficient estimation.
 In Simulation (b), the two methods are compared when the error distribution is skewed similarly to skew-$t$ but not exactly skew-$t$. 
  In Figure \ref{figure1} provides diagnostics to check whether the skew-$t$ error assumption is valid.
In Figure \ref{figure1} (A), 
the red curve of the true error distribution and the blue dashed curve of a skew-$t$ distribution are drawn over the histogram of generated error from a mixture of skew-$t$ and gamma distributions.
Although the two distributions are not in the same distribution family, they have very similar shapes.
In Figure \ref{figure1} (B), the skew-$t$ Q-Q plot of errors is represented. 
Similarly, although the errors were generated from the perturbed skew-$t$ distribution, the errors appear to follow a skew-$t$ distribution in the Q-Q plot.
Figure \ref{figure1} (C) shows the histogram of residuals obtained from the linear regression using \texttt{selm}.
Here, the estimated skew-$t$ distribution was drawn as a blue dashed curve over the histogram with the true error distribution shown as a red curve. 
\citet{arellano2013centred} denoted the parameters $(\xi,\omega^2,\alpha)$ as direct parameters (DP).
In \citet{azzalini2013skew}, it is noted that squares of scaled DP residuals follow a  $\chi^2$ distribution.
The  \textsf{R} package \texttt{sn} provides a P-P plot diagnostics for the fitted model.
Figure \ref{figure1} (D) displays the $\chi^2$ P-P plot of scaled DP residuals obtained from using  \texttt{selm}.
Since most points lie close to a straight line, the skew-$t$ error assumption is appropriate.
In Table \ref{table:SELM}, the results show that both estimators are consistent and that the proposed semiparametric estimator has a smaller standard error than estimator obtained following \citet{azzalini2020some}.
Thus, the proposed semiparametric method is useful when it is difficult to assume an exact distribution of errors.

\subsection{Real data example}

I analyzed a dataset of 202  Australian athletes, which includes 13 variables that reflect athletes' physical characteristics, such as body mass index.
The dataset `ais' can be downloaded from the \textsf{R} package \texttt{sn}. 
The following linear model is considered for the dataset.
\bse
Y_i=\beta_0+\beta_1 X_{1i}+\beta_2 X_{2i}+\epsilon_i,
\ese
for $i=1,2,\cdots,202$, where $Y_{i}$ is the body fat percentage for $i$th athlete,
and $X_{1i}$ and $X_{2i}$ are the body mass index and lean body mass, respectively, for the $i$th athlete. 
The estimation can be conducted according to the following procedure.
\begin{enumerate}[(a)]
\item Calculate the OLS estimator. Check the residuals.
\item If OLS is not appropriate, use the proposed semiparametric estimation method.
\item From the residual pattern obtained from (b), find a specific parametric distribution for errors. Conduct the parametric estimation using the assumed distribution.
\end{enumerate}

\begin{figure}
\includegraphics[scale=0.75]{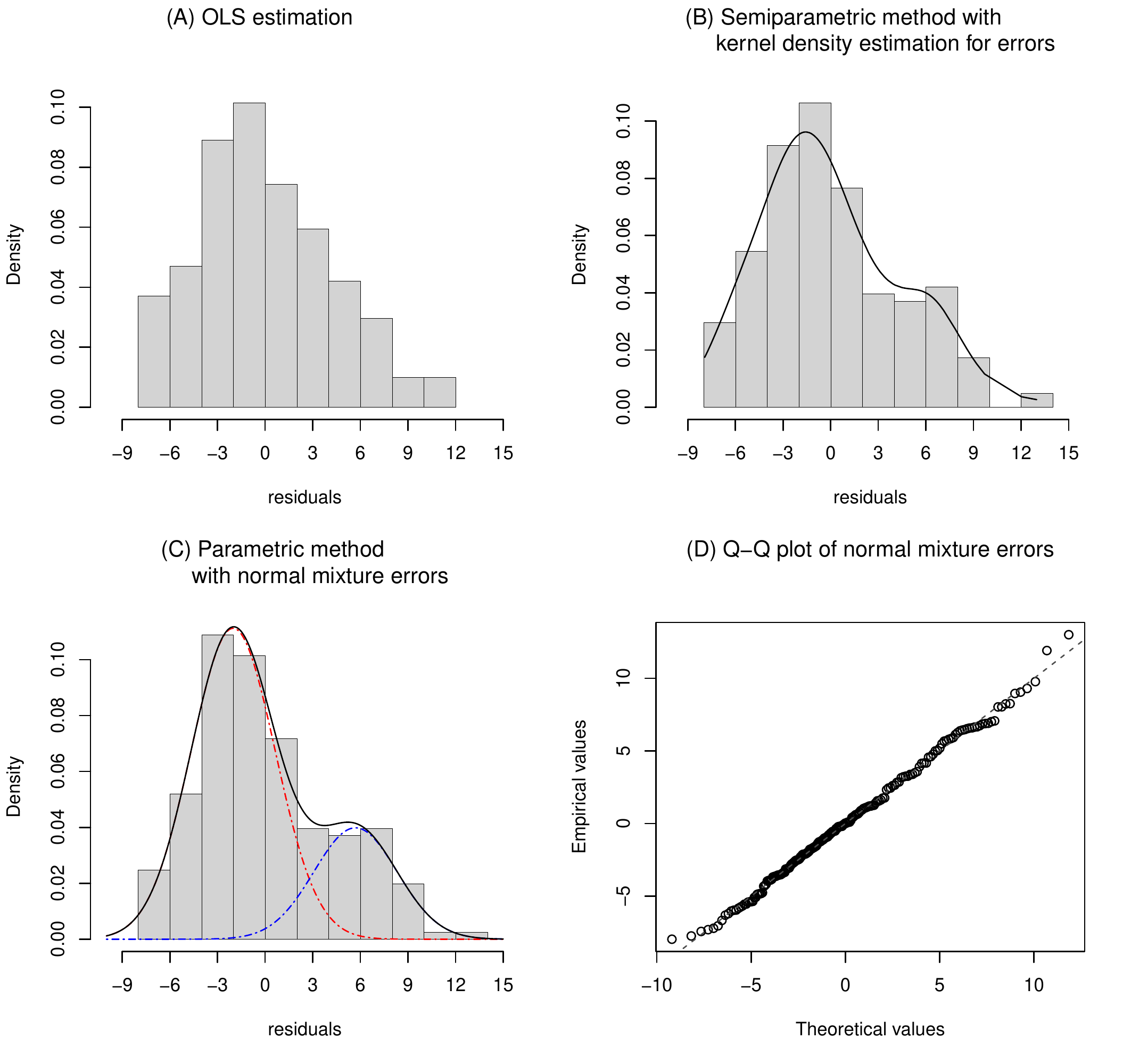}
\caption{ Diagnostics for the three methods. 
(A) The histogram of residuals of OLSE. 
(B) The histogram of residuals obtained from the semiparametric method with kernel density estimation for errors.
The estimated kernel density for residuals is overlayed.
(C) The histogram of residuals obtained from the parametric method with normal mixture errors.
The black curve represents a mixture of two normal densities. One of the normal distributions is shown as a red dashed curve and the other as a blue dashed curve.
(D) Q-Q plot of the estimated normal mixture distribution.  \label{Diagnostics_ais}
}
\end{figure}

We can check the diagnostics for the three methods in Figure \ref{Diagnostics_ais}.
In Figure \ref{Diagnostics_ais} (A),  the histogram of residuals of OLS estimation does not look symmetric.
The Shapiro-Wilk test statistic is 0.9811 with a $p$-value of 0.0081, which also supports that the normal error assumption is not valid.
Next, in order to identify the shape of the errors, a semiparametric method with kernel density estimation for errors was used. 
Figure \ref{Diagnostics_ais} (B) represents the histogram of residuals obtained from the method.
Since the residuals have a bimodal shape in Figure \ref{Diagnostics_ais} (B), 
it is natural to assume a mixture of two normal distributions for errors.
 Accordingly, we can modify the  efficient function  (\ref{eq:seff}) and solve it.
 In Figure \ref{Diagnostics_ais} (D), the Q-Q plot of the estimated normal mixture distribution is displayed. 
 The points are closely located on the line, which implies that the  normal mixture distribution assumption is appropriate for errors.
The results are shown in Table \ref{table:ais_result}.
The residual variances are 17.77, 18.55 and 18.39 for the (a) OLS estimation, (b) semiparametric method with kernel density and (c) parametric method with normal mixture, respectively.
Next, we need to select a better model between (b) and (c).
In Table \ref{simu3}, $n=200$ is not enough in terms of the 95\% coverage when using kernel density estimation for errors to estimate three parameters of interests.
Because   the number of observations is not sufficient for the kernel density estimation for errors  and Figure \ref{Diagnostics_ais} (C) and (D) support that the normal mixture assumption for errors is valid,
it would be better to select (b)  for the final model.

\section{Discussion}

In this paper, I have derived a semiparametric efficient score function for a homoscedastic regression model without any distribution assumption of errors based on \citet{tsiatis2006semiparametric}.
The estimated variance of the estimator reaches the asymptotic efficiency bound. 
Although the proposed method is superior in efficiency,  the method is not always available.
It is not available in a regression model containing only the intercept because there exists no derivative function of the intercept. 
We can modify the error-related terms of the derived semiparametric efficient function using parametric assumptions for errors or by kernel density estimation for errors. Although a nonparametric method can be used without error distribution assumptions, it may reduce precision when the number of samples is not sufficient. When we have a sparse dataset, only the parametric estimation approach for errors can be applicable. However, the parametric error model can lead to inaccurate results if the error assumptions are incorrect. Since neither the parametric approach nor the nonparametric approach can be said to be absolutely superior, it is necessary to use both methods appropriately according to the situation. Even if the nonparametric method is not finally selected because of the small number of sample sizes, an approximate shape of the error density can be detected using the nonparametric error estimation method in the intermediate process of estimation. In this regard, a nonparametric error estimation is useful in helping us find an appropriate parametric error density. When the number of samples is sufficient, in particular the shape of the error is skewed, the nonparametric estimation for errors can be highly useful.

In Section \ref{sec:simu_nonlinear_regression}, it is verified that nonlinear regression with two coefficients and $n=300$ samples is sufficient for nonparametric estimation of errors through the simulation. Since this is just one example, more simulations should be conducted to study the number of samples sufficient to use nonparametric methods for various situations in future work. In addition, it may be difficult to calculate the estimated variance if the matrix of the efficiency bound is close to singular. Solving the singularity issue can also be an interesting future study topic. 
 
\section*{Funding}
This research was supported by  a National Research Foundation of Korea (NRF) grant funded by the Korean Government (NRF-2020R1F1A1A01074157).






\bibliographystyle{apa}
\bibliography{regression}

\section*{Supporting Information}
Additional information for this article is available.

\newpage

\begin{table}
\caption{Results of simulation (b) with the parameters $(\beta_1,\beta_2, v)\trans=(12.0,-0.5, 3.7)\trans$. Model errors were generated from a Gumbel distribution. 
For the parametric approaches,  (1) Gumbel (true) pdf and (2) normal pdf were used. Additionally, (3) kernel density estimation for errors was used.  }  
 \label{simu2}
{\scriptsize
\begin{tabular}{cc|rrrr|rrrr|rrrr} \hline
	&		& \multicolumn{4}{c|}{(1)  True pdf }  & \multicolumn{4}{c|}{(2) Normal pdf}   &\multicolumn{4}{c}{(3) Kernel density}  \\ 
$n$	&	Parameter	&	Estimate	&	SE1	&	SE2	&	95\%cvg	&	Estimate	&	SE1	&	SE2	&	95\%cvg		&	Estimate	&	SE1	&	SE2	&	95\%cvg		\\ \hline
\multirow{3}{*}{$200$} 	&	$\beta_1$ 	&	11.9691 	&	{\bf 0.6339} 	&	0.5825 	&	96.8 	&	12.0347 	&	0.7956 	&	0.7581 	&	93.9 	&	11.9769 	&	0.6688 	&	0.6257 	&	98.3 	\\
	&	$\beta_2$	&	-0.4976 	&	{\bf 0.0311} 	&	0.0306 	&	95.1 	&	-0.4999 	&	0.0371 	&	0.0365 	&	95.6 	&	-0.4976 	&	0.0328 	&	0.0321 	&	95.1 	\\
	&	$ v$	&	3.6258 	&	0.5051 	&	0.4665 	&	98.2 	&	3.6172 	&	0.5211 	&	0.4819 	&	98.1 	&	3.6183 	&	{\bf 0.5042} 	&	0.4714 	&	98.3 	\\ \hline
																											
\multirow{3}{*}{$300$} 	&	$\beta_1$ 	&	11.9900 	&	{\bf 0.5214} 	&	0.4853 	&	94.8 	&	12.0168 	&	0.6720 	&	0.6325 	&	97.7 	&	12.0212 	&	0.5510 	&	0.5164 	&	95.2 	\\
	&	$\beta_2$	&	-0.5002 	&	{\bf 0.0272} 	&	0.0253 	&	95.9 	&	-0.5013 	&	0.0328 	&	0.0301 	&	96.5 	&	-0.5014 	&	0.0285 	&	0.0264 	&	96.6 	\\
	&	$ v$	&	3.6684 	&	{\bf 0.4192} 	&	0.3968 	&	92.3 	&	3.6649 	&	0.4276 	&	0.4073 	&	92.4 	&	3.6631 	&	0.4196 	&	0.3994 	&	92.6 	\\ \hline
																											
\multirow{3}{*}{$500$} 	&	$\beta_1$ 	&	11.9887 	&	{\bf 0.3984} 	&	0.3835 	&	95.7 	&	11.9988 	&	0.5065 	&	0.4982 	&	95.0 	&	11.9832 	&	0.4093 	&	0.3992 	&	95.9 	\\
	&	$\beta_2$	&	-0.5004 	&	{\bf 0.0203} 	&	0.0197 	&	94.2 	&	-0.5004 	&	0.0236 	&	0.0235 	&	93.6 	&	-0.4998 	&	0.0205 	&	0.0203 	&	93.1 	\\
	&	$ v$	&	3.6688 	&	0.3226 	&	0.3145 	&	97.4 	&	3.6801 	&	0.3296 	&	0.3227 	&	98.0 	&	3.6715 	&	{\bf 0.3223} 	&	0.3154 	&	97.6 	\\\hline
																											
\multirow{3}{*}{$1000$} 	&	$\beta_1$ 	&	12.0054 	&	{\bf 0.2755} 	&	0.2745 	&	95.9 	&	12.0219 	&	0.3626 	&	0.3543 	&	94.6 	&	12.0060 	&	0.2870	&	0.2843	&	94.7 	\\
	&	$\beta_2$	&	-0.5002 	&	{\bf 0.0141} 	&	0.0140 	&	94.8 	&	-0.5006 	&	0.0170 	&	0.0167 	&	93.7 	&	-0.5006 	&	0.0144	&	0.0143	&	95.3 	\\
	&	$ v$	&	3.6837	&	0.2345	&	0.2279	&	96.1 	&	3.6899 	&	0.2425 	&	0.2336 	&	95.7 	&	3.6826 	&	{\bf 0.2340}	&	0.2278	&	95.9 	\\ \hline
\end{tabular}
}
\end{table}

\begin{table}
\caption{Results of simulation (c) with the parameters $(\beta_1,\beta_2, v)\trans=(12.0,-0.5, 6.4)\trans$. Model errors were generated from a mixture of two Gaussian distributions. 
For the parametric approaches, (1) Gaussian mixture (true) pdf and (2) normal pdf were used. Additionally, (3) kernel density estimation for errors was used. }  
\label{simu3}
{\scriptsize
\begin{tabular}{cc|rrrr|rrrr|rrrr} \hline
	&		& \multicolumn{4}{c|}{(1) True pdf}  & \multicolumn{4}{c|}{(2) Normal pdf}   &\multicolumn{4}{c}{(3) Kernel density}  \\ 
$n$	&	Parameter	&	Estimate	&	SE1	&	SE2	&	95\%cvg	&	Estimate	&	SE1	&	SE2	&	95\%cvg		&	Estimate	&	SE1	&	SE2	&	95\%cvg		\\ \hline
\multirow{3}{*}{$200$} 	&	$\beta_1$ 	&	12.0036 	&	{\bf0.2717} 	&	0.2582 	&	94.4 	&	12.0933 	&	1.0913 	&	1.0438 	&	92.8 	&	11.9986 	&	0.2867 	&	0.2590 	&	93.5 	\\	
	&	$\beta_2$	&	-0.4992 	&	{\bf0.0197} 	&	0.0193 	&	95.6 	&	-0.5012 	&	0.0497 	&	0.0491 	&	89.9 	&	-0.4992 	&	0.0223 	&	0.0219 	&	96.1 	\\	
	&	$ v$	&	6.3770 	&	{\bf0.2442} 	&	0.2507 	&	94.5 	&	6.3576 	&	0.2935 	&	0.2943 	&	93.6 	&	6.3798 	&	0.2601 	&	0.2576 	&	94.4 	\\		  \hline													
\multirow{3}{*}{$300$} 	&	$\beta_1$ 	&	11.9992 	&	{\bf 0.2118} 	&	0.2110 	&	94.3 	&	12.0167 	&	0.8671 	&	0.8539 	&	96.7 	&	12.0075 	&	0.2212 	&	0.2124 	&	94.4 	\\	
	&	$\beta_2$	&	-0.5003 	&	{\bf0.0159} 	&	0.0158 	&	95.5 	&	-0.4987 	&	0.0411 	&	0.0400 	&	92.0 	&	-0.5000 	&	0.0176 	&	0.0176 	&	95.9 	\\	
	&	$ v$	&	6.3962 	&	{\bf0.2005} 	&	0.2052 	&	94.9 	&	6.3889 	&	0.2336 	&	0.2407 	&	97.4 	&	6.4003 	&	0.2090 	&	0.2102 	&	95.8 	\\	\hline			%
\multirow{3}{*}{$500$} 	&	$\beta_1$ 	&	12.0016 	&	{\bf0.1499} 	&	0.1655 	&	95.9 	&	11.9879 	&	0.6789 	&	0.6598 	&	95.2 	&	11.9992 	&	0.1603 	&	0.1666 	&	95.4 	\\	
	&	$\beta_2$	&	-0.4999 	&	{\bf0.0119} 	&	0.0123 	&	95.6 	&	-0.4987 	&	0.0318 	&	0.0311 	&	95.0 	&	-0.4994 	&	0.0131 	&	0.0133 	&	95.6 	\\	
	&	$ v$	&	6.4013 	&	{\bf0.1511} 	&	0.1588 	&	96.3 	&	6.3894 	&	0.1846 	&	0.1856 	&	94.5 	&	6.4026 	&	0.1615 	&	0.1620 	&	96.0 	\\	\hline

\multirow{3}{*}{$1000$} 	&	$\beta_1$ 	&	11.9998 	&	{\bf0.1130} 	&	0.1177 	&	95.9 	&	11.9975 	&	0.4586 	&	0.4694 	&	96.0 	&	11.9970 	&	0.1213	&	0.1184	&	95.6 	\\	
	&	$\beta_2$	&	-0.5000 	&	{\bf0.0085} 	&	0.0088 	&	94.9 	&	-0.5000 	&	0.0214 	&	0.0221 	&	95.4 	&	-0.5000 	&	0.0092	&	0.0092	&	95.3 	\\	
	&	$ v$	&	6.4029	&	{\bf0.1058}	&	0.1123	&	95.2 	&	6.3933 	&	0.1288 	&	0.1312 	&	94.5 	&	6.3979 	&	0.1131	&	0.1138	&	94.7 	\\	 \hline
\end{tabular}
}
\end{table}

\begin{table}
\begin{center}
\caption{Results of the Shapiro‒Wilk test when normal errors are assumed for simulations (a) and (b). The corresponding numbers are the percentage of cases in which residuals violate the assumption of normality.} \label{table:shapiro}
\begin{tabular}{c|r|c} \hline
$n$	&  (a) Gumbel errors& (b) Gaussian mixture errors\\ \hline
$200$ &	99.8\% &100.0\% \\
$300$ & 	100.0\% & 100.0\%\\
$500$ & 	 100.0\% &100.0\%	\\ 
$1000$ &  100.0\% &100.0\%	\\	\hline
\end{tabular}
\end{center}
\end{table}

\begin{table}
\begin{center}
\caption{Results of MLE and the semiparametric estimation  for the data with skew-$t$ errors and perturbed skew-$t$ errors. \label{table:SELM} } 
\begin{tabular}{lc|rrrr|rrrr} \hline
	&		& \multicolumn{4}{c|}{MLE}  & \multicolumn{4}{c}{Semiparametric estimation}    \\
	&	Parameter	&	Estimate	&	SE1	&	SE2 & 95\%cvg	&	Estimate	&	SE1	&	SE2  &95\%cvg\\  \hline
\multirow{4}{*}{100\% Skew-$t$} 	&	$\beta_0=5.00$ 	&	4.9815 	&	{\bf 0.3573} 	&	0.3438 	&	93.7	&	4.9825 	&	0.3686 	&	0.3506 	&	94.4 	\\
&	$\beta_1=1.00$ 	&	1.0047 	&	{\bf 0.0962} 	&	0.0948 	&	94.9	&	1.0079 	&	0.0980 	&	0.0961 	&	95.4 	\\
&	$\beta_2=1.80$	&	1.7938 	&	{\bf 0.1653} 	&	0.1586 	&	94.4	&	1.7943 	&	0.1691 	&	0.1618 	&	95.1 	\\
&	$ v=5.19$	&		&		&		&		&	5.1301 	&	0.6169 	&	0.5724 	&	93.5 	\\  \hline
\multirow{4}{*}{\parbox{3cm}{70\% Skew-$t$ \\ 30\% Gamma}}  &	$\beta_0=5.00$ 	&	4.9793 	&	0.4674 	&	0.4854 	&	95.3	&	4.9964 	&	{\bf0.4638} 	&	0.4844 	&	96.0 	\\
&	$\beta_1=1.00$ 	&	0.9974 	&	0.1339 	&	0.1347 	&	95.1	&	0.9989 	&	{\bf0.1310} 	&	0.1323 	&	97.5 	\\
&	$\beta_2=1.80$	&	1.7944 	&	0.2215 	&	0.2256 	&	95.7	&	1.7929 	&	{\bf0.2194} 	&	0.2223 	&	95.6 	\\
&	$ v=10.39$	&		&		&		&		&	9.7122 	&	1.2439 	&	1.2846 	&	96.4 	\\ \hline
\end{tabular}
\end{center}
\end{table}

\begin{table}
\begin{center}
\caption{Results of regression analysis for the three methods. \label{table:ais_result}}
\begin{tabular}{c|p{1.4cm}p{1.9cm}|p{1.7cm}p{1.7cm}|p{1.7cm}p{1.7cm}} \hline
	&	\multicolumn{2}{c|}{(a) OLS estimation}	&	\multicolumn{2}{c|}{(b) Semiparametric method}	& \multicolumn{2}{c}{(c) Parametric method } 	\\	
	&			&				&\multicolumn{2}{c|}{ with kernel density }	&  \multicolumn{2}{c}{with normal mixture }   \\ 
Parameter	&	\multicolumn{1}{r}{Estimate}	&	\multicolumn{1}{c|}{SE}	&\multicolumn{1}{r}{Estimate}	&	\multicolumn{1}{c|}{SE}	&	\multicolumn{1}{r}{Estimate}	&	\multicolumn{1}{c}{SE}	\\ \hline
$\beta_0$ 	&	\multicolumn{1}{r}{-0.5439}	&	\multicolumn{1}{c|}{2.4350}  &	\multicolumn{1}{r}{0.0143} 	&	\multicolumn{1}{c|}{1.7211} 	&	\multicolumn{1}{r}{-1.1146}	&	\multicolumn{1}{c}{1.7526} 	\\
$\beta_1$ 	&	\multicolumn{1}{r}{1.9650}	&	\multicolumn{1}{c|}{0.1490}  &	\multicolumn{1}{r}{1.6966} 	&	\multicolumn{1}{c|}{0.1594} 	&	\multicolumn{1}{r}{1.7224} 	&	\multicolumn{1}{c}{0.1154} 	\\
$\beta_2$	&	\multicolumn{1}{r}{-0.4787} 	&	\multicolumn{1}{c|}{0.0326}  &	\multicolumn{1}{r}{-0.3924} 	&	\multicolumn{1}{c|}{0.0488} 	&	\multicolumn{1}{r}{-0.3841}	&	\multicolumn{1}{c}{0.0279} 	\\
$ v$	&		&		&\multicolumn{1}{r}{18.3009} 	&	\multicolumn{1}{c|}{1.7166} 	&	\multicolumn{1}{r}{18.4599 }	&	\multicolumn{1}{c}{1.6070} 	\\ \hline
\end{tabular}
\end{center}
\end{table}

\end{document}